\documentclass[preprint,aps]{revtex4}

\usepackage{graphicx}
\usepackage{dcolumn}
\usepackage{bm}
\usepackage{url}

\begin{document}

\preprint{APS/123-QED}

\title{Spectroscopy of Neutron-Rich $^{168,170}$Dy:\\Yrast Band Evolution Close to the $N_{\mathrm{p}}N_{\mathrm{n}}$ Valence Maximum}

\author{P.-A.~S\"{o}derstr\"{o}m}\affiliation{Department of Physics and Astronomy, Uppsala University, SE-75121 Uppsala, Sweden}
\author{J.~Nyberg}\affiliation{Department of Physics and Astronomy, Uppsala University, SE-75121 Uppsala, Sweden}
\author{P.~H. Regan}\affiliation{Department of Physics, University of Surrey, Guildford, GU2 7XH, UK}
\author{A.~Algora}\affiliation{IFIC, CSIC-Univ. Valencia, Apartado Oficial 22085, 46071 Valencia, Spain}
\author{G.~de~Angelis}\affiliation{Istituto Nazionale di Fisica Nucleare, Laboratori Nazionali di Legnaro, I-35020 Legnaro, Italy}
\author{S.~F.~Ashley}\affiliation{Department of Physics, University of Surrey, Guildford, GU2 7XH, UK}
\author{S.~Aydin}\affiliation{Dipartimento di Fisica dell'Universit\`{a} and INFN, Sezione di Padova, I-35122 Padova, Italy}
\author{D.~Bazzacco}\affiliation{Dipartimento di Fisica dell'Universit\`{a} and INFN, Sezione di Padova, I-35122 Padova, Italy}
\author{R.~J.~Casperson}\affiliation{Wright Nuclear Structure Laboratory, Yale University, New Haven, Connecticut 06520, USA}
\author{W.~N.~Catford}\affiliation{Department of Physics, University of Surrey, Guildford, GU2 7XH, UK}
\author{J.~Cederk\"{a}ll}\affiliation{PH Department, CERN 1211, Geneva 23, Switzerland}
\author{R.~Chapman}\affiliation{School of Engineering and Science, University of the West of Scotland, Paisley PA1 2BE, Scotland, UK}
\author{L.~Corradi}\affiliation{Istituto Nazionale di Fisica Nucleare, Laboratori Nazionali di Legnaro, I-35020 Legnaro, Italy}
\author{C.~Fahlander}\affiliation{Department of Physics, Lund University, SE-22100 Lund, Sweden}
\author{E.~Farnea}\affiliation{Dipartimento di Fisica dell'Universit\`{a} and INFN, Sezione di Padova, I-35122 Padova, Italy}
\author{E.~Fioretto}\affiliation{Istituto Nazionale di Fisica Nucleare, Laboratori Nazionali di Legnaro, I-35020 Legnaro, Italy}
\author{S.~J.~Freeman}\affiliation{Schuster Laboratory, University of Manchester, Manchester M13 9PL, UK}
\author{A.~Gadea}\affiliation{IFIC, CSIC-Univ. Valencia, Apartado Oficial 22085, 46071 Valencia, Spain}\affiliation{Istituto Nazionale di Fisica Nucleare, Laboratori Nazionali di Legnaro, I-35020 Legnaro, Italy}
\author{W.~Gelletly}\affiliation{Department of Physics, University of Surrey, Guildford, GU2 7XH, UK}
\author{A.~Gottardo}\affiliation{Istituto Nazionale di Fisica Nucleare, Laboratori Nazionali di Legnaro, I-35020 Legnaro, Italy}
\author{E.~Grodner}\affiliation{Istituto Nazionale di Fisica Nucleare, Laboratori Nazionali di Legnaro, I-35020 Legnaro, Italy}
\author{C.~Y.~He}\affiliation{Istituto Nazionale di Fisica Nucleare, Laboratori Nazionali di Legnaro, I-35020 Legnaro, Italy}
\author{G.~A.~Jones}\affiliation{Department of Physics, University of Surrey, Guildford, GU2 7XH, UK}
\author{K.~Keyes}\affiliation{School of Engineering and Science, University of the West of Scotland, Paisley PA1 2BE, Scotland, UK}
\author{M.~Labiche}\affiliation{School of Engineering and Science, University of the West of Scotland, Paisley PA1 2BE, Scotland, UK}
\author{X.~Liang}\affiliation{School of Engineering and Science, University of the West of Scotland, Paisley PA1 2BE, Scotland, UK}
\author{Z.~Liu}\affiliation{Department of Physics, University of Surrey, Guildford, GU2 7XH, UK}
\author{S.~Lunardi}\affiliation{Dipartimento di Fisica dell'Universit\`{a} and INFN, Sezione di Padova, I-35122 Padova, Italy}
\author{N.~M\u{a}rginean}\affiliation{Istituto Nazionale di Fisica Nucleare, Laboratori Nazionali di Legnaro, I-35020 Legnaro, Italy}\affiliation{National Institute for Physics and Nuclear Engineering, RO-77125, Bucharest-Magurele, Romania}
\author{P.~Mason}\affiliation{Dipartimento di Fisica dell'Universit\`{a} and INFN, Sezione di Padova, I-35122 Padova, Italy}
\author{R.~Menegazzo}\affiliation{Dipartimento di Fisica dell'Universit\`{a} and INFN, Sezione di Padova, I-35122 Padova, Italy}
\author{D.~Mengoni}\affiliation{Dipartimento di Fisica dell'Universit\`{a} and INFN, Sezione di Padova, I-35122 Padova, Italy}
\author{G.~Montagnoli}\affiliation{Dipartimento di Fisica dell'Universit\`{a} and INFN, Sezione di Padova, I-35122 Padova, Italy}
\author{D.~Napoli}\affiliation{Istituto Nazionale di Fisica Nucleare, Laboratori Nazionali di Legnaro, I-35020 Legnaro, Italy}
\author{J.~Ollier}\affiliation{STFC Daresbury Laboratory, Daresbury, Warrington, WA4 4AD, UK}
\author{S.~Pietri}\affiliation{Department of Physics, University of Surrey, Guildford, GU2 7XH, UK}
\author{Zs.~Podoly\'{a}k}\affiliation{Department of Physics, University of Surrey, Guildford, GU2 7XH, UK}
\author{G.~Pollarolo}\affiliation{Dipartimento di Fisica Teorica, Università di Torino, and Istituto Nazionale di Fisica Nucleare, I-10125 Torino, Italy}
\author{F.~Recchia}\affiliation{Istituto Nazionale di Fisica Nucleare, Laboratori Nazionali di Legnaro, I-35020 Legnaro, Italy}
\author{E.~\c{S}ahin}\affiliation{Istituto Nazionale di Fisica Nucleare, Laboratori Nazionali di Legnaro, I-35020 Legnaro, Italy}
\author{F.~Scarlassara}\affiliation{Dipartimento di Fisica dell'Universit\`{a} and INFN, Sezione di Padova, I-35122 Padova, Italy}
\author{R.~Silvestri}\affiliation{Istituto Nazionale di Fisica Nucleare, Laboratori Nazionali di Legnaro, I-35020 Legnaro, Italy}
\author{J.~F.~Smith}\affiliation{School of Engineering and Science, University of the West of Scotland, Paisley PA1 2BE, Scotland, UK}
\author{K.-M.~Spohr}\affiliation{School of Engineering and Science, University of the West of Scotland, Paisley PA1 2BE, Scotland, UK}
\author{S.~J.~Steer}\affiliation{Department of Physics, University of Surrey, Guildford, GU2 7XH, UK}
\author{A.~M.~Stefanini}\affiliation{Istituto Nazionale di Fisica Nucleare, Laboratori Nazionali di Legnaro, I-35020 Legnaro, Italy}
\author{S.~Szilner}\affiliation{Ruder Boskovic Institute, HR 10 001, Zagreb, Croatia}
\author{N.~J.~Thompson}\affiliation{Department of Physics, University of Surrey, Guildford, GU2 7XH, UK}
\author{G.~M.~Tveten}\affiliation{PH Department, CERN 1211, Geneva 23, Switzerland}\affiliation{Department of Physics, University of Oslo, Oslo, Norway}
\author{C.~A.~Ur}\affiliation{Dipartimento di Fisica dell'Universit\`{a} and INFN, Sezione di Padova, I-35122 Padova, Italy}
\author{J.~J.~Valiente-Dob\'{o}n}\affiliation{Istituto Nazionale di Fisica Nucleare, Laboratori Nazionali di Legnaro, I-35020 Legnaro, Italy}
\author{V.~Werner}\affiliation{Wright Nuclear Structure Laboratory, Yale University, New Haven, Connecticut 06520, USA}
\author{S.~J.~Williams}\affiliation{Department of Physics, University of Surrey, Guildford, GU2 7XH, UK}
\author{F.~R.~Xu}\affiliation{School of Physics and State Key Laboratory of Nuclear Physics and Technology, Peking University, Beijing 100871, People's Republic of China}
\author{J.~Y.~Zhu}\affiliation{School of Physics and State Key Laboratory of Nuclear Physics and Technology, Peking University, Beijing 100871, People's Republic of China}

\date{\today}

\begin{abstract}
The yrast sequence of the neutron-rich dysprosium isotope $^{168}$Dy has been studied using multi-nucleon transfer reactions following tcollisions between a 460-MeV $^{82}$Se beam and a $^{170}$Er target. The reaction products were identified using the PRISMA magnetic spectrometer and the $\gamma$ rays detected using the CLARA HPGe-detector array. The $2^{+}$ and $4^{+}$ members of the previously measured ground state rotational band of $^{168}$Dy have been confirmed and the yrast band extended up to $10^{+}$. A tentative candidate for the $4^{+}\to2^{+}$ transition in $^{170}$Dy was also identified. The data on these nuclei and on the lighter even-even dysprosium isotopes are interpreted in terms of Total Routhian Surface calculations and the evolution of collectivity in the vicinity of the proton-neutron valence product maximum is discussed.
\end{abstract}

\pacs{21.10.Re, 27.70.+q, 23.20.Lv}
\maketitle


Our microscopic understanding of nuclei rests to a large extent upon the well-known shell model with the magic neutron and proton numbers occurring near to stability at $N,Z=2$, 8, 20, 28, 50, 82 and 126. The features associated with this model appear most clearly for nuclei in the vicinity of closed shells. Another important approach to the nuclear many-body problem is the macroscopic understanding which is based on collective properties of nuclei. These properties are most prominent in the regions around the doubly mid-shell nuclei, with large numbers of both valence protons and neutrons which maximizes the number of possible neutron and proton interactions. The importance of the number of proton-neutron interactions, which is equal to the product of valence nucleons $N_{\mathrm{p}}N_{\mathrm{n}}$, for quadrupole collectivity is well known \cite{casten2}. It has been shown that both the energy, $E(2^{+})$, and the reduced transition probability, $B(\mathrm{E}2)$, of the first $2^{+}$ state, as well as the energy ratio $E(4^{+})/E(2^{+})$ have a smooth dependence on this quantity \cite{casten0,casten1,mach,zhao}. 

\begin{figure*}[ht]
\includegraphics[angle=-90,width=0.75\columnwidth]{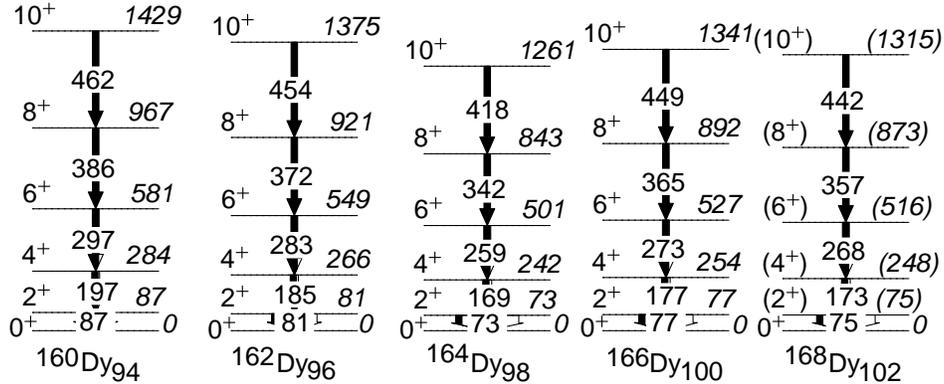}
\caption{Ground state rotational bands for dysprosium isotopes with $N=94-102$ from \protect\cite{160dy_band, 162dy_band,164dy_band,166dy_band,168dy} and the current work for $6^+$--$10^+$ in $^{168}$Dy.\label{fig:bands}}
\end{figure*}

Neglecting any potential sub-shell closures, the nucleus with the largest number of valence particles with $A<208$ is $^{170}_{\ 66}$Dy$_{104}$. Accordingly it should be one of the most collective of all nuclei, in its ground state \cite{paddy}. However, at present nothing is known experimentally about $^{170}$Dy, which makes $^{168}_{\ 66}$Dy$_{102}$ the nucleus with the largest $N_{\mathrm{p}}N_{\mathrm{n}}$ value below $^{208}$Pb with excited states reported in the current literature \cite{168dy}. It is also the most neutron-rich, even-$N$ dysprosium isotope that has been studied to date. The isotope $^{169}$Dy has been identified but no excited states have been observed \cite{PhysRevC.42.R1171}. Looking how $E(2^{+})$ changes in Fig.~\ref{fig:bands}, the dysprosium isotopes appear to become more collective, i.e. have lower $E(2^{+})$ values, with increasing neutron numbers from $^{160}$Dy up to $^{164}$Dy \cite{160dy_band, 162dy_band,164dy_band}. At $^{166}$Dy, however, $E(2^{+})$ increases again \cite{166dy,166dy_band}, suggesting that the maximum collectivity in dysprosium isotopes occurs at $N=98$ instead of at $N=104$. A maximum in the collectivity at $N=104$ might be expected since the neighbouring even-$Z$ elements above dysprosium (i.e. Er, Yb and Hf) have a minimum of their $2^{+}$ state energy at midshell ($N=104$) \cite{168dy}. The only spectroscopic measurement published on $^{168}$Dy to date is from a $\beta$-decay experiment \cite{168dy} and preliminary results of the present experiment published in \cite{170dy_lnl}. These results show a decrease of $E(2^{+})$ and $E(4^{+})$ for $^{168}$Dy compared to $^{166}$Dy. The data show an irregular lowering of $\gamma$-ray energies and level energies at both $N=98$ ($^{164}_{\ 66}$Dy$_{98}$) and $N=102$ ($^{168}_{\ 66}$Dy$_{102}$) compared to their nearest neighbors. Furthermore, it has been suggested that $^{170}$Dy could be the single best case in the entire Segr\'{e} chart for the empirical realization of the SU(3) dynamical symmetry \cite{PhysRevC.31.1991} and therefore spectroscopic information on this nucleus and its near neighbours is valuable in testing the effectiveness of the interacting boson approximation for such nuclei. 

Due to the neutron-rich nature of $^{168}$Dy it is not possible to study this nucleus and its neighbours using  traditional methods of high-spin spectroscopy that employ fusion-evaporation reactions. To populate states in nuclei with $A>164$ in the dysprosium isotopic chain, isotope separation on-line followed by $\beta$-decay measurements \cite{168dy}, in-beam fragmentation \cite{zsolt} and deep inelastic multi-nucleon transfer reactions together with a binary partner gating technique have been used so far. However, the nucleus $^{170}$Dy is very hard to study even with these techniques. The last technique is the one used in the current work. For a recent review on deep inelastic multi-nucleon transfer reactions, see ref.~\cite{transfer}.

The nuclei studied in this article were populated using multi-nucleon transfer reactions of a $^{82}$Se beam and a 500-$\mu$g/cm$^{2}$ thick self-supporting $^{170}$Er target. The primary $^{82}$Se beam was delivered by the Tandem XTU-ALPI accelerator complex at LNL \cite{piave} and had an energy of 460~MeV with a typical intensity of $\sim$2~pnA. Beam-like fragments were identified event-by-event using the PRISMA magnetic spectrometer \cite{prisma}. PRISMA was placed at the grazing angle of $52^{\circ}$. The energies of $\gamma$ rays from both the beam-like and target-like fragments were measured using the CLARA $\gamma$-ray detector array \cite{clara}.

\begin{figure}[ht]
 \includegraphics[width=0.45\columnwidth]{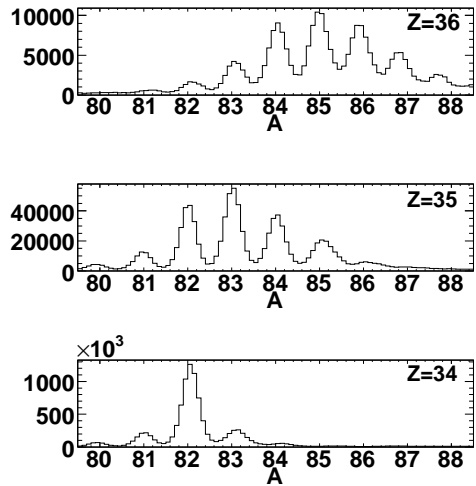}
\caption{One-dimensional projections of the mass distributions of the beam-like fragments as identified by PRISMA. \label{label_fig_mass}}
\end{figure}

The PRISMA magnetic spectrometer consists of a 50~cm length and 30~cm diameter quadrupole magnet and a dipole magnet with 1.2-m radius of curvature; it covers a solid angle of 80~msr. The atomic number ($Z$) resolution in this experiment was $Z/\Delta Z \approx 65$ and the mass resolution was $A/\Delta A \approx 200$ for elastic scattering of $^{82}$Se. At the entrance of PRISMA, 25~cm from the target, a position-sensitive micro-channel plate (MCP) that measures the $(x,y)$ position and the time of the ion entering PRISMA \cite{prisma_mcp} was placed. After the magnets, a 1-m wide multiwire parallel-plate avalanche counter (MWPPAC) segmented in ten elements that measured the $(x,y)$ position and gave a time reference for the ion at the end of the spectrometer was mounted. This was
followed by an ionization chamber segmented into four sections
along the optical axis of PRISMA and ten sections transverse to it
which measure the energy and energy-loss characteristics of the
transmitted heavy ion \cite{prisma2}. From the energy measurements in the ionization chambers the atomic number $Z$ of the ion could be determined using $\Delta E-E$ techniques. By reconstructing the trajectory in PRISMA from the position measurements in the MCP and the MWPPAC and the time-of-flight (TOF), the mass of the ion was determined \cite{szilner}. In Fig.~\ref{label_fig_mass} the atomic number ($Z$) and mass ($A$) distribution of the beam-like fragments is shown for $Z=34-36$ (i.e. Se, Br and Kr ions). The velocity vector of the beam-like fragments were obtained from the $(x,y)$ position in the MCP and the TOF between the MCP and the MWPPAC. Table~\ref{table:relative} gives the relative experimental yields for $Z=36$ ions.

\begin{table} [ht] 
\caption{\label{table:relative}Yields for the beam-like fragments (BLF) relative to $^{84}$Kr and the corresponding target-like fragments (TLF) as obtained from PRISMA in coincidence with CLARA, $Y_{\mathrm{rel}}$. Note that the experimental yields are without any acceptance or transmission corrections for the ions through the PRISMA spectrometer and that the TLF yields is an upper limit since neutron evaporation is not taken into account.}
  \begin{tabular*}{0.5\columnwidth}{@{\extracolsep{\fill}}llcc}
    \hline
    BLF& TLF$_{\mathrm{max}}$& Transfer & $Y_{\mathrm{rel}}$ \\
    \hline
    $^{81}$Kr & $^{171}$Dy & +2p-3n & $0.038\pm0.007$ \\
    $^{82}$Kr & $^{170}$Dy & +2p-2n & $0.159\pm0.009$ \\
    $^{83}$Kr & $^{169}$Dy & +2p-1n & $0.457\pm0.016$ \\
    $^{84}$Kr & $^{168}$Dy & +2p    & $1.000\pm0.029$ \\
    $^{85}$Kr & $^{167}$Dy & +2p+1n & $1.230\pm0.035$ \\
    $^{86}$Kr & $^{166}$Dy & +2p+2n & $1.084\pm0.035$ \\
    $^{87}$Kr & $^{165}$Dy & +2p+3n & $0.652\pm0.028$ \\
    $^{88}$Kr & $^{164}$Dy & +2p+4n & $0.306\pm0.022$ \\
    $^{89}$Kr & $^{163}$Dy & +2p+5n & $0.154\pm0.016$ \\
    $^{90}$Kr & $^{162}$Dy & +2p+6n & $0.069\pm0.010$ \\
    \hline
  \end{tabular*}
\end{table}

In its full complement, CLARA consists of 25 Compton-suppressed CLOVER detectors (in this experiment 23 CLOVER detectors were mounted) distributed in a hemisphere opposite to the entrance of PRISMA, covering the angles $104^{\circ}$--$256^{\circ}$ with respect to the entrance direction of the ions in the spectrometer. Each CLOVER detector is in turn composed of four germanium crystals surrounded by a BGO Compton-suppression shield. The triggers used in the experiment were coincidences between the MCP and CLARA or the MWPPAC and CLARA. For an event to be considered valid the ion had to be detected in the MCP, the MWPPAC and the ionization chamber, but not in any of the side ionization chambers. There also had to be at least one coincident $\gamma$ ray detected in CLARA.

Doppler correction was performed event-by-event using the velocity vectors measured by PRISMA. This gave an energy resolution of 4.4~keV (0.7~\%) at 655~keV for the beam-like fragments and 5.8~keV (1.1~\%) at 542~keV for the target-like fragments. The velocity vector of the target-like fragment was obtained using simple two-body kinematics between the beam-like fragment and the unobserved binary reaction partner (the target-like fragment before neutron evaporation). Since the PRISMA MCP has an angular resolution of $\Delta \theta < 1^{\circ}$ the Doppler broadening of the beam-like fragments is mainly due to the finite angular size of the CLARA crystals.

\begin{figure}[ht]
\includegraphics[width=0.5\columnwidth]{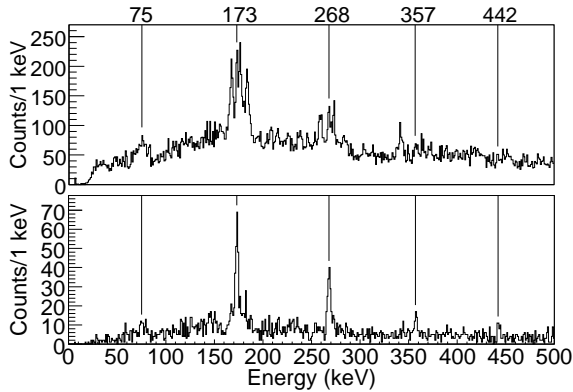}
\caption{\label{fig:evaporation}Spectrum of $\gamma$-ray energies from target-like fragments gated on the beam-like fragments $^{84}$Kr (top) and on beam-like fragments $^{84}$Kr plus a short time-of-flight (bottom). The transitions identified as the rotational band in $^{168}$Dy are marked with solid lines.
}
\end{figure}
\begin{figure}[ht]
\includegraphics[width=0.45\columnwidth]{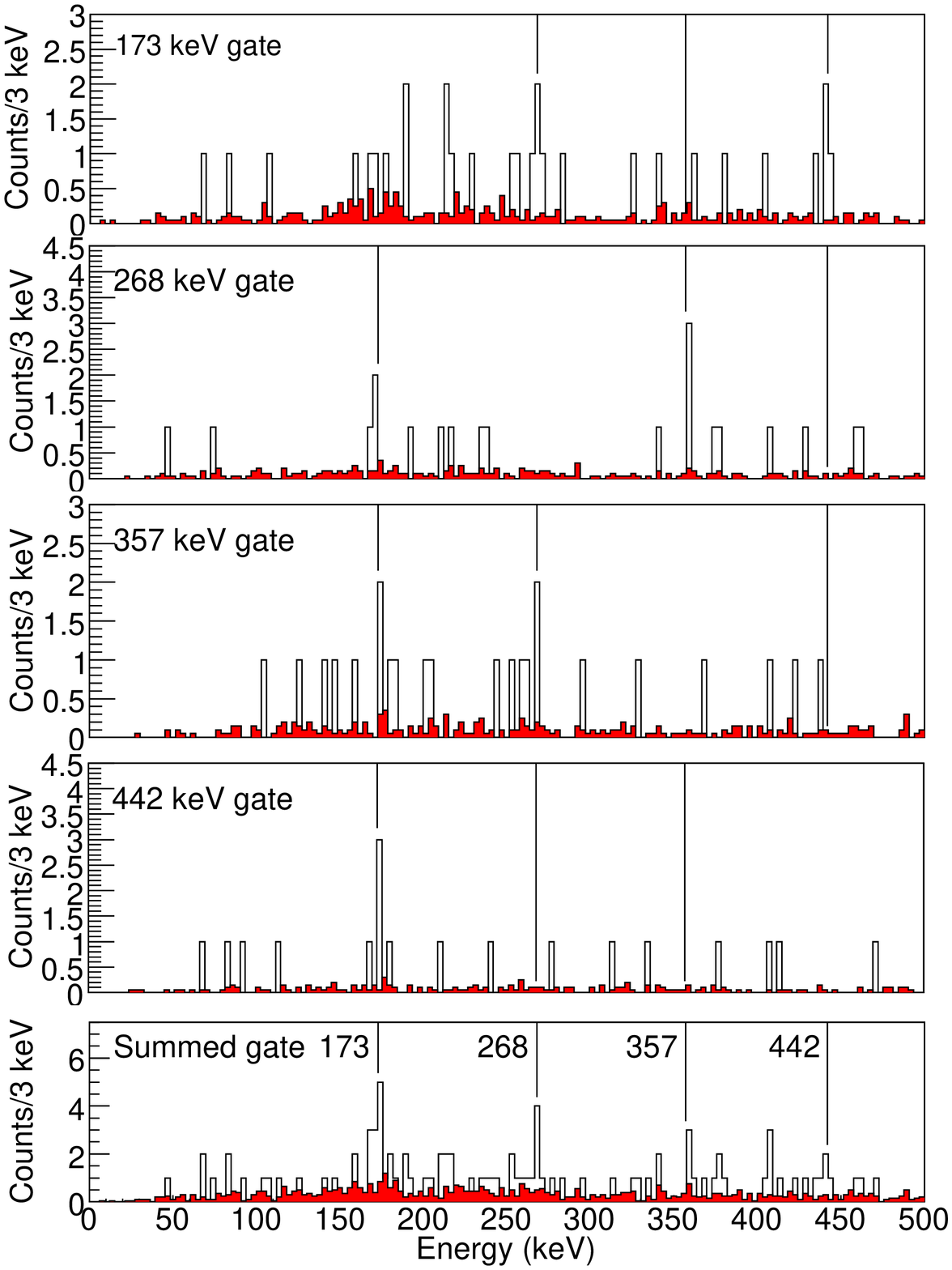}
\caption{\label{fig:gg}(Color online) Coincidence $\gamma$-ray spectra gated on the beam-like fragments $^{84}$Kr and on the $\gamma$-ray energies 173~keV, 268~keV, 357~keV, 442~keV and the summed spectrum (unfilled histogram) from top to bottom. An estimation of the background using adjacent gates are also shown (filled histogram). The background gates are about 20 times the width of the gates on the $\gamma$-ray peaks and normalized relative to the size of the peak gates. The transitions identified as the rotational band in $^{168}$Dy are marked with solid lines.}
\end{figure}

Using the measured $Z$ of the beam-like fragments, the atomic number of the target-like fragments was adopted under the assumption that there was no evaporation of charged particles. Using the same procedure an upper limit on the mass of the target-like fragment was obtained as the assumption of no
evaporated particles is violated, particular by neutron emission. Since only an upper limit of the mass was obtained from the PRISMA information, the $\gamma$-ray spectra not only contained lines from the maximum-mass dysprosium binary partner, but also from lighter dysprosium isotopes associated with neutron evaporation channels. The target-like fragments could thus not be uniquely identified event-by-event. It was, however, possible to suppress the contribution from the lighter dysprosium isotopes by using the TOF information, corresponding to the total kinetic energy loss \cite{valiente}. Since an energy greater than the separation energy of a neutron needs to be transferred from the beam-like fragment in order for the neutron to evaporate, beam-like fragments with a partner that evaporates neutrons will, on average, have a lower velocity than beam-like fragments originating from the pure binary transfer reaction channels. Requiring a high velocity of the beam-like fragment by setting conditions on the TOF information from PRISMA, the peaks corresponding to fragments that have undergone neutron evaporation could be heavily suppressed, see Fig.~\ref{fig:evaporation}. The two previously reported $\gamma$-ray transitions in $^{168}$Dy at 75~keV and 173~keV are clearly apparent in this spectrum. Three previously unreported transitions at 268~keV, 357~keV and 442~keV are also identified in the spectrum. The efficiency and internal-conversion corrected relative intensities of the $\gamma$ rays shown in Fig.~\ref{fig:evaporation} are $1.7\pm0.5$, $1.00\pm0.12$, $0.62\pm0.09$, $0.30\pm0.07$ and $0.25\pm0.08$, respectively, assuming that the transitions are of E2 character.
To verify that these transitions originate from the same decay sequence, the $\gamma\gamma$-coincidence method was applied to the data. The results from the $\gamma\gamma$-coincidence analysis (using an $A$ and $Z$ selection) are shown in Fig.~\ref{fig:gg}, which shows that the 173, 268, 357 and 442~keV transitions form a mutually coincident $\gamma$-ray cascade, assumed to be the ground-state band excitations in $^{168}$Dy. The $2^+\to0^+$ transition is heavily converted and very close in energy to the corresponding transitions in the other even-$N$ dysprosium isotopes and thus not included in the $\gamma\gamma$-coincidence analysis. The three new $\gamma$ rays are assigned on the basis of systematics to be the transitions associated with the yrast $6^{+}\to4^{+}$, $8^{+}\to6^{+}$ and $10^{+}\to8^{+}$ decays in $^{168}$Dy. The level scheme deduced for $^{168}$Dy from the current work is shown in Fig.~\ref{fig:bands}.

\begin{figure}[ht]
\includegraphics[width=0.5\columnwidth]{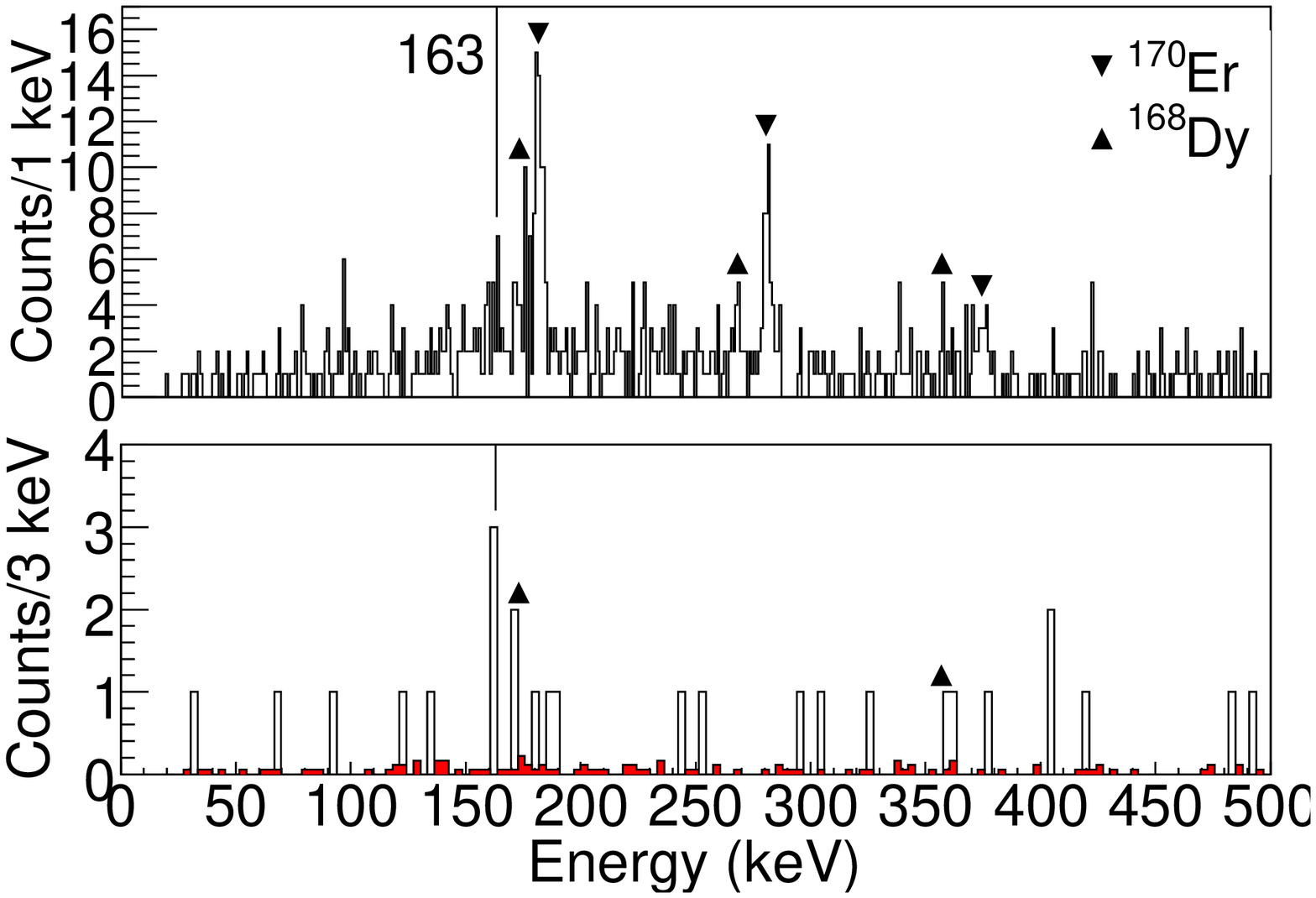}
\caption{\label{fig:dy170}(Color online) Spectrum of $\gamma$-ray energies from target-like fragments gated on the beam-like fragments $^{82}$Kr plus time-of-flight (top). Coincidence $\gamma$-ray spectra gated on the beam-like fragments $^{82}$Kr, time-of-flight plus the $\gamma$-ray energy 777~keV in the beam-like fragments (bottom). An estimation of the background using adjacent gates is also shown (filled histogram). The background gates are about 20 times the width of the gates on the $\gamma$-ray peaks and normalized relative to the size of the peak gates. The tentative $\gamma$ ray associated with the yrast $4^{+}\to2^{+}$ transition in $^{170}$Dy is marked with a solid line.}
\end{figure}

Since no reported $\gamma$-ray lines exist in $^{170}$Dy which can be used for $\gamma\gamma$-coincidences, the $\gamma$-ray energy of 777~keV, associated with the $2^{+}\to0^{+}$ transition in $^{82}$Kr \cite{82kr}, which is the binary reaction partner of $^{170}$Dy, was used. By using gates on the beam-like fragments, $\gamma$ rays from neutron evaporation channels from the respective target-like fragments can be TOF suppressed in the $\gamma\gamma$-coincident spectrum. In Fig.~\ref{fig:dy170}, both the singles spectrum and the $\gamma\gamma$-coincidence spectrum (using an $A$, $Z$ and TOF selection) are shown. Due to the finite $Z$ resolution there is a large leakage from $^{170}$Er in the singles spectrum. Both in the singles spectrum and the $\gamma\gamma$-coincident spectrum a peak appears at 163~keV. This peak is tentatively identified from the dysprosium energy systematics as a candidate for the transition associated with the yrast $4^{+}\to2^{+}$ decay in $^{170}$Dy. The corresponding $2^+\to0^+$ $\gamma$-ray would be too weak to be observed because of internal conversion and detector efficiency.

\begin{figure}[ht]
\includegraphics[width=0.9\textwidth]{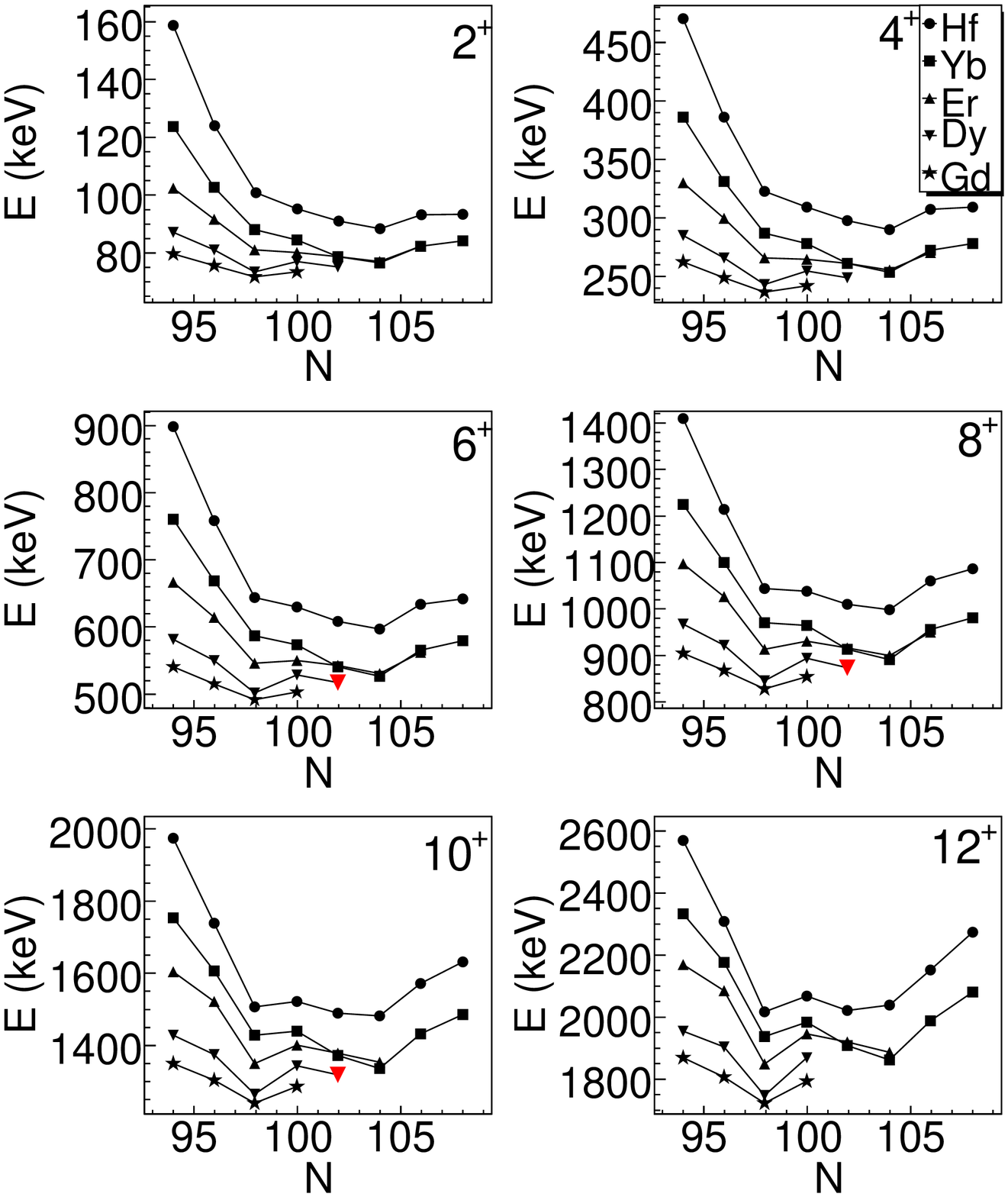}
\caption{\label{fig:dysyst}(Color online) Experimental energy levels for $Z=64$ (gadolinium), $Z=66$ (dysprosium), $Z=68$ (erbium), $Z=70$ (ytterbium) and $Z=72$ (hafnium) isotopes with $N=94-108$. The experimental values are obtained from \protect\cite{166dy_band,A168,A170,A172,A174,A176,A178,A180,164dy_band,162dy_band,PhysRevC.70.014313,174Er,160dy_band,168dy,A158}
 and the current work (red triangles online).}
\end{figure}

As reported in Ref.~\cite{168dy} an irregularity in the energy systematics of the yrast $2^{+}$ and $4^{+}$ states exists at $N=98$ for $Z=64$ (gadolinium) and $Z=66$  (dysprosium). Extending the systematics to higher spin shows that this irregularity also appears further up in the yrast band of $Z=66$, showing that this is a systematic effect and not only a small fluctuation at low energies, see fig.~ \ref{fig:dysyst}. This irregularity also appears in elements with larger $Z$ at higher spin. According to existing data, the global energy minimum at $N=104$ is clear at low spins and stays quite stable up to $I^{\pi} = 12^{+}$. 
However, for $Z=68$ (erbium) the energy levels of the isotopes with $N=102$ and $N=104$ increase relative to $N=98$, even above the corresponding energy levels in $Z=70$ (ytterbium), causing $N=98$ to become a new global minimum. The data on $^{168}_{\ 66}$Dy$_{102}$ presented in this paper shows no such increase relative to $^{164}_{\ 66}$Dy$_{98}$.

\begin{figure}[ht]
\includegraphics[width=0.9\textwidth]{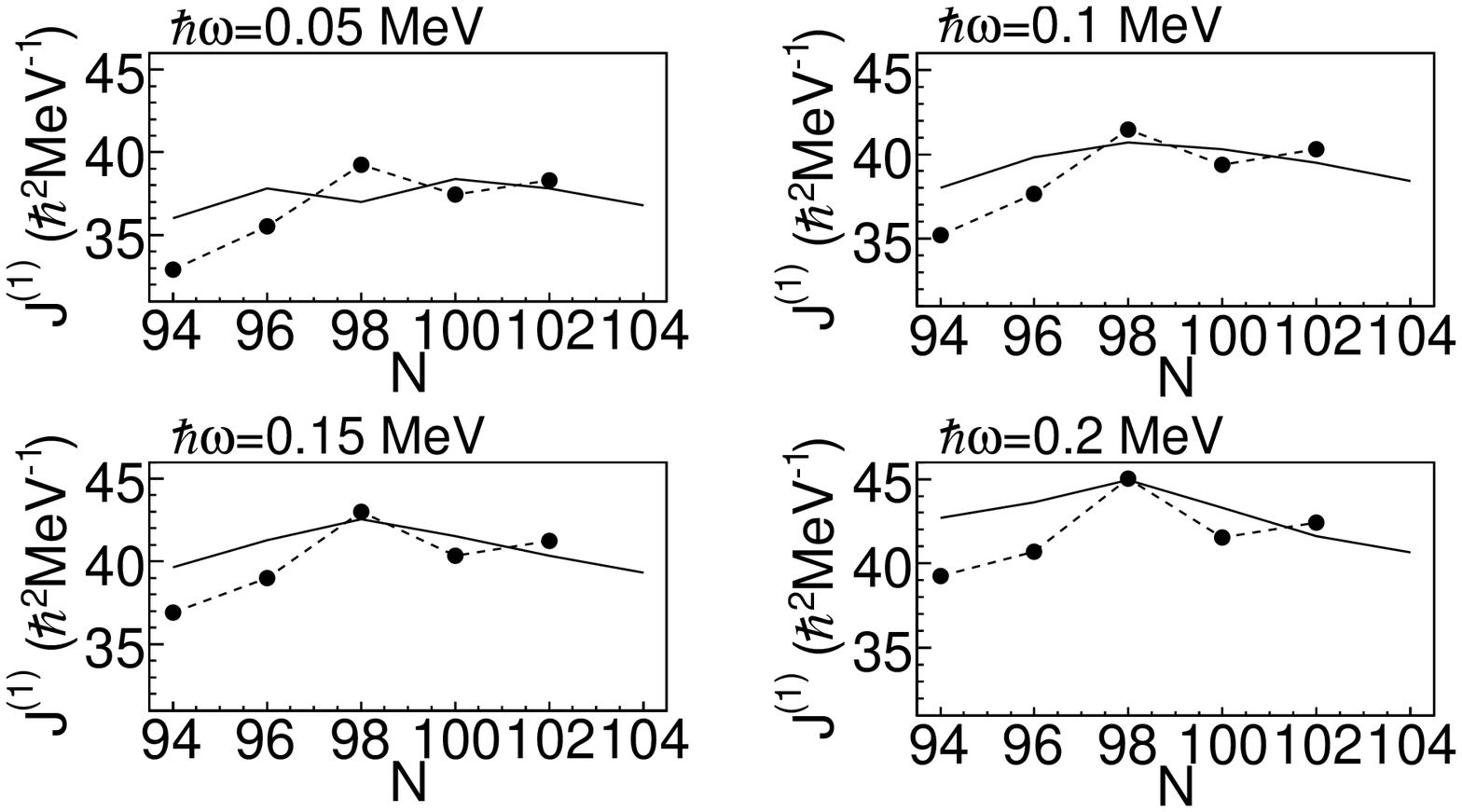}
\caption{\label{fig:dytrs}Experimental (circles) and calculated (solid line) moments of inertia at a rotational frequency of $\hbar\omega=0.05$, $0.10$, $0.15$ and $0.20$~MeV for dysprosium isotopes with $N=94-104$. The experimental values are obtained using linear interpolation between measured rotational frequencies from \protect\cite{160dy_band, 162dy_band,164dy_band,166dy_band,168dy} and the current work.}
\end{figure}

The irregularity at $N=98$ is not reproduced by the results of the Total Routhian Surface calculations \cite{trs1,trs2,trs3} shown in Fig.~\ref{fig:dytrs}. The irregularity could be caused by a strong interaction between the ground-state band and the two quasi-neutron band in $^{164}_{\ 66}$Dy$_{98}$ \cite{neils}. The interpretation that the irregularity is an effect in $^{164}_{\ 66}$Dy$_{98}$ and not in neighbouring isotopes is strengthened by the tentative identification of the $4^{+}\to2^{+}$ transition at 163~keV in $^{170}$Dy as well as higher spin systematics in neighbouring elements. The energy systematic of the yrast band of $^{168}_{\ 66}$Dy$_{102}$ as well as the tentative identification of the $4^{+}\to2^{+}$ transition at 163~keV in $^{170}$Dy further suggests that  maximum collectivity in dysprosium isotopes does not occur at $N=98$. 

The current work demonstrates the possibility of identifying heavy, high-$Z$, target-like fragments and their velocity vectors from kinematic reconstruction using information about the lighter beam-like fragments. We have also shown that it is possible to suppress heavily the effects of neutron evaporation in the analysis of this kind of experiment. This allows the use of binary partner distributions to get information on high-spin states in neutron-rich, deformed rare-earth nuclei inaccessible in traditional fusion-evaporation reactions. We have extended the
level scheme of $^{168}$Dy to $10\hbar$ and proposed a tentative energy
for the $4^{+}\to2^{+}$ transition in $^{170}$Dy. 

\begin{acknowledgments}
This work was partially supported by the European Commission within the Sixth Framework Programme through I3-EURONS contract RII3-CT-2004-506065, the Swedish Research Council, EPSRC/STFC (UK) and U.S. DOE grant No. DE-FG02-91ER40609.
\end{acknowledgments}

\clearpage 

\end{document}